\documentclass[12pt,preprint]{aastex}

\shorttitle{Equivalence for astrophyscal jets}
\shortauthors{Mendoza et al.}

\usepackage{amsmath,amssymb,amsbsy,amsfonts}
\usepackage{amssymb}
\usepackage[british]{babel}

\title{A possible hydrodynamical equivalence for astrophysical jets}
\author{S. Mendoza, X. Hernandez \& W. H. Lee }
\affil{Instituto de Astronom\'{\i}a, Universidad Nacional 
       Aut\'onoma de M\'exico, AP 70-264, Distrito Federal 04510,
       M\'exico}
\email{sergio@astroscu.unam.mx}

\begin{abstract}
  The idea of a unified model for quasars and \(\mu\)--quasars has
been considered for a long time, despite the different environments
and physical conditions where both classes of objects reside.  Here we
show the existence of a simple scaling law, relating the maximum size
of a jet to the properties of the gas medium into which it expands.
This appears to be valid for all types of hydrodynamical jets, and can
be thought of as a broad unified model.  The expansion velocity of the jet
and the physical properties of the surrounding gas combine in such a
way that a limit to the maximum extent of jets at different scales can
be obtained.
\end{abstract}


\begin{document}

\maketitle

\section{Introduction}
\label{introduction}

  When astrophysical jets are formed, they begin to expand into
their surrounding medium.  A jet is capable of this expansion because
the outflow, produced by the central engine, generates a ram pressure
capable of elongating it.  At the head of a particular jet, the kinetic
energy of the particles is thermalised through a shock wave or \emph{hot
spot}.  After a particular fluid particle crosses the hot spot, it is
recycled back to the neighbourhood of the jet itself.  This generates a
cavity or \emph{cocoon} that, together with the jet, expands through the
environment.   The cocoon keeps the jet collimated so that its transverse
length does not grow too much as it expands.  This standard picture for
the expansion of an extragalactic (hydrodynamical) radio jet was first
proposed by \citet{scheuer74}.

  Today, we know that jets are not only associated with quasars, but
that they occur at a range of different astrophysical scales. Some jets
are observed in recently formed star systems.  These are associated
with Herbig--Haro objects and are commonly named Herbig--Haro jets
\citep{reipurth01}.  Jets can also form in galactic stellar binary systems
\citep{mirabel94}, and are so similar to their quasar extragalactic
counterparts that they were termed \( \mu \)--quasar jets.  
Finally, observations of gamma ray burst afterglows (of the
long variety) in distant galaxies, have led recently to the idea that
jets are associated with these sources (see \citet{meszaros02} for a review).

  Aside from a difference of \( 6 - 7 \) orders of magnitude covered
by the maximum size of jets, their kinetic power output varies
depending on the nature of the jet itself.  Observations show that
some extragalactic, and all Herbig--Haro jets, expand at
non--relativistic velocities.  On the other hand, the fluid particles
inside quasar, \(\mu\)--quasar and \( \gamma \)--ray burst jets expand
at relativistic velocities.  Scheuer's~(1974) simple hydrodynamical
model has been used as the basic model for all kind of jets that we
observe in the universe.

  Using dimensional analysis arguments, we propose in this article that there
is a simple way to relate the maximum size of an astrophysical jet
with the properties of its surrounding environment. The result appears
to be valid both for low--velocity jets in the Newtonian regime and for
highly relativistic outflows such as those found in \( \gamma \)--ray bursts.

  Dimensional analysis techniques have been applied to the study
of powerful jets in radio galaxies interacting with their external
environment by \citet{falle91} and \citet{kaiser97}.  However, the nature
of their models show a self--similar indefinite growth of cocoon and jet,
possibly due  to the omission of gravity from the forces acting on the
growing cocoon, since they concentrated their analysis on the initial
growth phase.

\section{Non--relativistic hydrodynamical jets}

  Let us first consider a non--relativistic jet, for example an AGN or
a Herbig--Haro object jet, that has a characteristic length \( r \).
The physical processes and the dimensionless combination of important
parameters that determine the dynamics of the expansion of the jet
are many and complicated \citep{begelman84,blandford90}.  For example,
internal shock waves that develop inside jets, radiative cooling, kinetic
power, magnetic fields, mass of the central engine and accretion power are
some of the important physical quantities that enter into the description
of an expanding jet. For the sake of simplicity, suppose that we do not
include magnetic fields on the description of the jet's dynamics and
consider a purely hydrodynamical jet.  Of all the possible dimensionless
combinations of physical parameters that describe the problem, one was
constructed by \citet{mendoza01}.  This parameter measures the strength
of the kinetic energy in the jet as compared to the gravitational energy
of the gas cloud where it is embedded. In order to build such parameter,
we proceed as follows: the gravity of the surrounding cloud is taken into
account by introducing Newton's gravitational constant \( G \), and the
physical properties of the cloud are determined by its average density
\( \rho \).  In general terms, this density does not only represent the
gas density of the surrounding material. It represents the \emph{total}
average density evaluated at a certain distance from the progenitor jet
source.  This means that the average density contains the gas density
plus the stellar mass density (for galactic jets) and dark matter mass
density (for extragalactic jets).  Finally, since the expanding jet is
roughly in pressure equilibrium with the surrounding cloud, the jet is
roughly characterised by its bulk velocity \( v \).

  If dissipative processes inside the jet (such as radiative cooling
and drag acting on the flow) are negligible then the four fundamental
quantities (\( r,\ G,\ \rho\ \text{and} \ v \)) are a good estimate of
the dynamical behaviour of the expanding jet in the cloud.  This assumes
that the expansion of the jet is close to ballistic and is of course a
first order approximation, dealing only with the purely hydrodynamical
aspect of the problem through dimensional analysis. One of the many
simplifications being made is to consider a constant density medium,
which is in all likelihood not true.  However, under the reasonable
assumption of a jet expanding from the centre of a spherically symmetric
region, or from an overdensity within a cloud, the average density is
well defined at each radius.

  For any mechanical problem there are three independent dimensions,
namely the dimensions of time \( t \), length \( l \), and mass \( m \).
Having restricted the problem to the above physics, Buckingham's theorem
of dimensional analysis \citep{sedov93} demands the existence of a unique
non--dimensional parameter \( \Lambda \) that will fully determine the
solution of the problem.  In the case of jets, this parameter is given by

\begin{equation}
 \Lambda  \equiv G \frac{  \rho r^2 }{ M^2 a^2 },
\label{eq.1}
\end{equation}

\noindent where \( M \equiv v / a \).  Formally this number is not the
Mach number of the jet flow since it is defined as the ratio of the jet
velocity to the sound speed \( a \) in the cloud \citep{mendoza01}.  These
authors showed that, for the particular problem they were discussing,
the dimensionless parameter \( \Lambda \) measured in units of \( M^{-2}
\) had the same value for jets formed in giant molecular clouds and for
the gaseous haloes of galaxies (the length \( r \) in their analysis was a
characteristic length for which deflection of jets by pressure gradients
might occur).  This equivalence hints at an underlying mechanism that
makes jets look the same at such different scales.

  As noted by \citet{mendoza01} the parameter \( \Lambda \) can also be
written as

\begin{equation}
  \Lambda = \frac{ 3 }{ 4 \pi } \frac{ G  \mathsf{M}(r)  / r }{ v^2 }
  \approx 1.2 \times 10^{-17} \left( \frac{ \mathsf{M}(r) }{ M_\odot } 
  \right) \left( \frac{ c }{ v } \right)^2 \left( \frac{ r }{ \textrm{kpc} } 
  \right)^{-1},
\label{eq.6}
\end{equation}

\noindent where \( { \mathsf{M}(r) } \) is the total mass within a sphere
of radius \( r \) and \( c \) is the speed of light.  The right hand
side of equation~\eqref{eq.6} is roughly the ratio of the gravitational
potential energy to the kinetic energy of a fluid element of the jet
evaluated at position \( r \). Formally, the total gravitational potential
energy acting on a fluid element on the jet needs to include not only
the gas, but also the central mass  and the dark matter (for the case
of extragalactic jets)  within radius \( r \). Let us consider from now
on that \( r \) represents the maximum length that a specific class of
jets can have. This maximum length is constrained by observations.

  If a particular jet is such that \( \Lambda \gg 1 \), then the total
gravity acting on the jet is so strong that the jet cannot expand away
from the central engine that is producing it.  On the other hand, when
\( \Lambda \ll 1 \) the jet can freely expand away from the progenitor
source, drilling a hole through the parent cloud.  Since the numerical
factor appearing in equation~\eqref{eq.6} is very small, it follows that
for any real jet, the parameter \( \Lambda \ll 1 \).  For example, if we
consider a Herbig--Haro jet, for which the total mass within a radius
\( r \) is a few \( M_\odot \) and the expansion velocity \( v \approx
10^{-3} c \), where \( c \) is the speed of light, in which typical
lengths  \( r \approx 1 \textrm{pc} \), we then obtain a value of \(
\Lambda \approx 10^{-7} \).   For the case of powerfull extragalactic
jets, as will be discussed later,  \( M(r) \approx 10^{15} \), \( r
\approx 1 \, \textrm{Mpc} \) and \( v \sim c \) which implies that \(
\Lambda \approx 10^{-5} \).  As will be shown below, in the case of a
relativistic jet we would replace \(\beta\) by \(\gamma \beta\), where
\( \beta = v / c \) and \( \gamma \) is the Lorentz factor of the flow.
This replacement reduces even more the value of \( \Lambda \).

  Since \( \Lambda \ll 1 \) for all types of jets (including
\(\mu\)--quasar and \( \gamma \)--ray burst jets), it would appear
that from merely considering gravitational potential and kinetic
energies, jets should still be capable of growing several orders of
magnitude in size beyond what is observed. Typical sizes of jets are
much smaller than what the energy balance of \(\Lambda \sim 1 \)
suggests. This indicates that the mechanism determining the maximum
extent of a jet must lie elsewhere.

  In view of this fact, let us return to equation~\eqref{eq.1} in its
original form, keeping the ratio \( v/a \) explicit, and performing
a re--scaling.  The maximum length that a Herbig--Haro jet can have is
\( r \approx 1\, \textrm{pc} \) \citep{reipurth01} and the surrounding
medium where those jets expand are cold molecular clouds for which
the particle number density \( n_\mathrm{H} \! \sim \!  {10}^{2} \,
\textrm{cm}^{-3} \) and the temperature \( T \!  \sim 10 \, \textrm{K}
\) \citep{spitzer98,hartmann98}.  We can thus write equation~\eqref{eq.6}
as

\begin{equation}
  \Lambda  \approx  \frac{ 10^ {-1} }{ M^2 } \left( \frac{ M }{
    10 \, \text{M}_\odot } \right)\left( \frac{ r }{ 1 \,
    \mathrm{pc} } \right)^{-1} \left( \frac{ T }{ 10 \, \mathrm{K} } 
    \right)^{-1}. 
\label{eq.2}
\end{equation}

\noindent  The total (stellar plus gas) mass \( \mathsf{M}(r) \) within a 
radius of \( \sim 1 \, \textrm{pc} \) is such that  \( 2 \lesssim 
\mathsf{M}(r) \lesssim 10 \text{M}_\odot \), so that the upper limit of
equation~\eqref{eq.2} is \( \Lambda \approx 10^{-1} / M^2 \).

  The largest extragalactic jets have typical lengths \( r \approx
2 \, \textrm{Mpc} \) \citep{ferrari98} and their surrounding
intergalactic medium is such that \( n_\mathrm{H} \approx {10}^{-4} \,
\textrm{cm}^{-3}, \) and \( T \approx {10}^{8} \, \textrm{K}
\) \citep{HEA2,galaxyformation}.  The total gas mass within this radius is
\( \mathsf{M}_\text{gas} \approx 10^{13} \, \text{M}_\odot \).  Clusters of
galaxies are such that the fraction of gas mass to dark matter is \( \sim
0.1 \) \citep[][and references within]{fabian02,fabian04}.  This means that the
total mass within a radius of \( \sim 2 \, \textrm{Mpc} \) is \(
\mathsf{M}(r) \approx 10^{14} \, \text{M}_\odot \).  With these values
equation~\eqref{eq.2} implies that \( \Lambda \approx 10^{-1} / M^2 \), the
same as the one obtained for Herbig--Haro jets.

  The fact that the numerical value of \( \Lambda \) in units of \(
M^{-2} \) is the same for Herbig--Haro jets and for non--relativistic
extragalactic jets appears to indicate that a certain underlying physical
mechanism is involved, fixing a maximum jet size.  In order to investigate
this, let us now rewrite equation~\eqref{eq.1} as

\begin{equation}
 \Lambda \approx  \frac{ 1 }{ M^2 } \frac{ r^2 }{ \lambda_\text{J}^2 },
\label{eq.3}
\end{equation}

\noindent where \( \lambda_\text{J} \approx a / \sqrt{ G \rho } \)
is the cloud's Jeans length.  Note that \( \lambda_\text{J} \) is the
scale--length for gravitational instability in a fluid having sound speed
\( a \) and subject to the gravity produced by a total density \( \rho
\), which includes all matter present. The above comes from setting \(
\rho^{-1}_\text{gas} \nabla p_\text{gas} = \nabla \phi  \), where \(
\phi \) is the gravitational potential.

Comparing  equation~\eqref{eq.3} with~\eqref{eq.2}
we find that if the relation \( \Lambda \approx 10^{-1}/M^{2} \) can be 
treated as universal, then

\begin{equation}
  \lambda_\text{J} \approx \sqrt{10} \, r.
\label{eq.4}
\end{equation}

\noindent One way of interpreting this general result is the following.
It appears that jets can grow inside gaseous clouds, up to the point
where they can no longer do so because a gravitational instability of
the surrounding material sets in.  In fact, the numerical factor of \(
\sim \sqrt{10} \) that appears in equation~\eqref{eq.4} can be thought
of as a form factor for the problem.  The Jeans stability criterion can
be applied to the gas inside the cocoon plus the swept up gas contained
within the outer shell surrounding it in the presence of an extra
central gravitational potential including dark matter or central star.
The relevant pressure \( p_\text{c} \) inside the cocoon decreases as \(
p_\text{c} \propto t^{-\alpha} \) with the expansion of the source in
time \( t \).  Here \( \alpha \approx 4/5 \, \text{--} \, 2 \) depending
on the density profile of the external gas \citep{kaiser97}.  Since the
gravitational force produced by the shocked gas inside the shell at
the boundaries of the cocoon remains the same and because the pressure
inside the cocoon tends to zero with the expansion of the source, there
might be a time when a gravitational instability on the ``ellipsoidal''
cocoon will develop, leading to a collapse of the shell.  In fact,
the regions closest to the head of the jet will not collapse because \(
\Lambda \ll 1 \).  However, regions of the shell closest to the central
engine will eventually collapse for sufficiently large cocoon sizes,
leading to the inflow of the external gas, which perhaps destroys the jet.

  There is however another physical interpretation that can be
given to equation~\eqref{eq.4}. \citet{alexander02} has proposed
that the important instability which affects the late evolution in
powerful FR--II radiogalaxies is the Rayleigh--Taylor instability
\citep{frieman54,chandrahydro}.  Due to the fact that Rayleigh--Taylor
instabilities grow proportional to \( \exp{ \left(4 \pi G \rho t \right)
} \) in time \( t \), they scale in exactly the same manner as the Jeans
length when the wavelength \( \lambda \) of the oscillations is such that
\( \lambda \gg \lambda_\text{J} \).  Once the gravitational force of the
swept up gas is of the same order of the cocoon pressure, the interface
between the swept up gas and the cocoon becomes Rayleigh--Taylor unstable
leading to the break in of the surrounding medium, most likely inhibiting
the growth of the jet.

  The mean size of powerful radio sources is \( \sim 300 \, \textrm{Kpc}
\) with maximum sizes reaching up to \( \sim 2 \, \textrm{Mpc} \).
In the scenario presented here, these mean sized objects are understood
as transient, being in the process of expansion. They will continue
to grow until they reach the limit given by equation~\eqref{eq.4} provided
the central engine remains active long enough. That no jets larger
than this limit have been found, and the otherwise extraordinary
coincidence of equation~\eqref{eq.4} at stellar and extragalactic scales
suggest that the physics behind these objects is the same. Of course,
other dimensionless parameters can be constructed, related to the many
other physical aspects of the problem, such as magnetic fields or nature
of the central engine. However, the numerical results of the previous
section suggest that the particular feature of the problem which we
are examining, i.e. maximum jet length, relates directly, or at least
primarily, to \(\Lambda\) through equation~\eqref{eq.4}.

\section{Relativistic Jets--Generalising Further}

  When a fully relativistic jet is introduced in the discussion, the
dimensional analysis is slightly more complex. This is because the
speed of light \( c \) has to be introduced as a fundamental parameter.
In this case, the quantity \( M \) is given by the ratio of the
proper velocity of the jet \( \gamma v \) to the proper velocity of
sound \( \gamma_a a \) in the medium \citep{chiu73,mendoza02},

\begin{equation}
  M = \frac{ \gamma v }{  \gamma_a a },
\label{eq.5}
\end{equation}

\noindent where \( \gamma = \left( 1 - v^2 / c^2 \right)^{-1/2} \) and \(
\gamma_a = \left( 1 - a^2 / c^2 \right)^{-1/2} \) represent the Lorentz
factors of the jet material and the sound speed respectively.  Since all
quantities, except the quantity \( M \), that appear in equation~\eqref{eq.1}
are measured in the proper frame of the gas, and because the sound
speed of the gas in the cloud is non--relativistic, it is clear that
the dimensionless parameter \( \Lambda \) given by equation~\eqref{eq.1}
also has the same form in the relativistic case.  The only difference
is that the number \( M \) is now defined by equation~\eqref{eq.5}.
Following the same procedure as before, it follows that equation~\eqref{eq.3}
should be valid in this situation as well.

  For the case of relativistic extragalactic jets, i.e. quasar jets, that
expand through an intergalactic medium similar to their non--relativistic
counterparts, it follows that \( \Lambda \sim 10^{-1} / M^2 \), so for
this particular case equation~\eqref{eq.4} is also valid.

  We can now repeat the calculations for the case of \( \mu \)--quasars
to test the generality of the result.  Most \( \mu \)--quasars show
jets with typical lengths \( r \approx 1 \textrm{pc} \) \citep[see][and
references therein]{corbel02}.  The total (stellar plus gas) mass within
\( 1 \, \textrm{pc} \) is such that \( \mathsf{M}(r) \lesssim 10 \,
\text{M}_\odot \). Since the temperature of the interstellar medium is
\( T \approx 10 \, \text{K} \) it follows that \( \mu \)--quasar jets
satisfy the condition given by equation~\eqref{eq.5}.

  As a final application in the extreme, consider cosmological \(
\gamma \)--ray bursts, GRBs \citep{fm95,meszaros02}.  The relevant
parameters are less well known than for the previous examples, but a
comparison is nevertheless possible. These events typically release \(
10^{51} \)~erg for a few seconds in a highly relativistic outflow, 
with Lorentz factors \( \gamma \approx  100-1000 \). The long variety,
for which X--ray, optical and radio counterparts have been observed
\citep{vkw00}, is probably related to the collapse of massive stars
\citep{woosley93,stanek03,hjorth03}. The outflow itself is apparently
highly collimated in many cases, manifesting itself as a relativistic
jet, and the physical size of the source has actually been measured
through radio scintillation, and is typically \( r \approx {10}^{-2} -
1 \, \textrm{pc} \) \citep{frail97,frail00}.  If the progenitors of GRBs
are massive stars, it is reasonable to assume that they will be found
in star--forming regions within the ISM. The density of the circumburst
environment has also been determined from broadband observations in
several cases, and ranges from \( 0.1 \, \textrm{cm}^{-3} \) to \( 100 \,
\textrm{cm}^{-3} \), with a canonical value of \( 10 \, \textrm{cm}^{-3}
\) \citep{pk02}. If the jet is directly interacting with the external ISM,
the temperature could be as low as \( 10 \, \textrm{K} \).  This gives
a value of \( \Lambda \) that does not violate our constraints, and
is lower than in the previous cases. If however, the jet is within the
wind--fed bubble produced by the massive progenitor star prior to the GRB,
the temperature will be much higher, perhaps \( 10^{6} \, \textrm{K} \)
\citep[see e.g.][]{chevalier04}. Inspection of equation~\eqref{eq.2}  shows
that this will decrease \( \Lambda \) even further, thus also satisfying
our proposed constraint that the right--hand side of equation~\eqref{eq.2} is
an upper stability limit for \( \Lambda \). One additional fact makes
GRB jets different from the other types considered here: the central
engine typically lasts for only a few tens of seconds for long--duration
bursts. It is most likely the case that this is what determines the
extent of the jet, but it nevertheless does so within the limits imposed
by equation~\eqref{eq.2} under the above interpretation.

\section{Discussion}

  The results presented in this article suggest that astrophysical jets at
all scales seem to obey the same physical law given by equation~\eqref{eq.4},
regardless of their surrounding environment or the nature of the jet
itself.  It is remarkable that a class of objects where the physics
is super relativistic and where the overall phenomena have a duration
of milli--seconds, over \( 15 \) orders of magnitude below that of
large radio galaxies, appear to satisfy (or at any rate not to deviate
significantly from) the predictions of equation~\eqref{eq.4}.

\begin{figure}
  \begin{center}
   \includegraphics[scale=1.0]{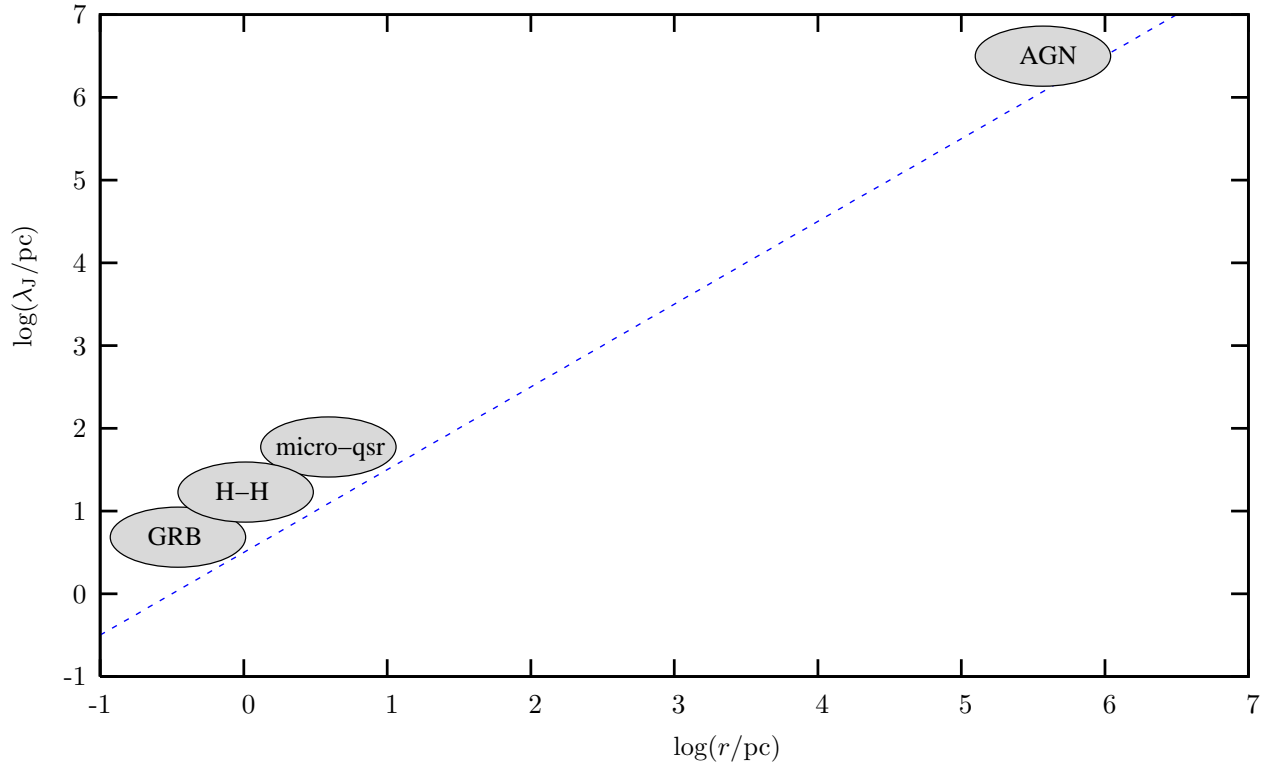}
  \end{center}
  \caption[Jeans length as a function of the jet radius]{ Hydrodynamical
	   jets are stable because there is a basic equilibrium
	   between the maximum length of the jet \( r \) and the Jeans
	   length \( \lambda_\text{J} \) of the surrounding medium
	   given by \( \lambda_\text{J} \sim \sqrt{10} r \).  The plot shows
	   this linear relationship for all known jets: \( \gamma \)--ray
	   burst (GRB) jets, Herbig--Haro (H--H) jets, \( \mu \)--quasar
	   (micro--qsr) jets and active galactic nuclei (AGN) jets. In
	   our analysis the region below the curve appears as a forbidden
	   region, with jets in the process of expansion or contraction
	   lying above it.}
\label{fig1}
\end{figure}

  The idea of unification of jets at all physical scales that
expand in very different environments has of course been considered
before. Much work has been made regarding the mechanism by which jets are
\emph{produced}, from \( \mu \)--quasar to AGN, with the magnetic field
usually playing an important role \citep{bp82,meier01,gc02,ppk03}.
Using the observed time variability, the scaling is in terms of
the mass of the central object, be it a black hole or a young star
(e.g. \citet{mirabel94}). In this article we have proposed a unification
scheme based on the physical properties between the expanding jet and 
its surroundings, described pictorically in Fig.~\ref{fig1}.  The straight
line is given by equation (1), and we have argued here that real jets must 
lie above it.

  Other physical causes of a jet's maximum size are possible, most
obviously the cut of an injection of particles and energy generated by
the central engine.  Also, for a purely hydrodynamical jet with a fixed
opening angle \( \theta \) that expands through an external gaseous
environment at constant pressure, lateral expansion of the jet becomes
important \citep{begelman84}.  As a constant momentum is distributed
over a growing area, ram pressure decreases, and a natural maximum
extent might be reached through pressure equilibrium.  Jet opening
angles have been studied and observed on many astrophysical sources,
mainly on extragalactic radio sources.  These observations show
globally well collimated jets over their large scale structure.  Indeed,
magnetic fields and internal shock waves \citep{begelman84,blandford90}
are probably important physical mechanisms that recollimate the jet
flow at large scales.  We hence believe that this explanation does not
account properly for the maximum length of jets.

  Also, well collimated self--similar jets and cocoons have been
modelled very successfully \citep{falle91,kaiser97}, albeit neglecting
self-gravity of the medium into which the jet expands, other than in
fixing the initial density profile of the surrounding medium.  This
very probably explains why no characteristic or limiting size for jets
appears in their analysis (their jets actually continue to expand
indefinitely). The analysis of the previous sections would indicate a
limiting size which is perhaps due to gravitational instabilities in 
the growing cocoon, which collapses onto the jet.  Most probably this
instability is a Rayleigh--Taylor instability arising from the
gravitational field produced by all matter within the cocoon
interacting with the swept up gas as the source expands
\citep{alexander02}.

 To conclude, we propose in this paper that purely hydrodynamical
jets of characteristic sizes are stable  because there is a fundamental
combination of definite physical quantities that limit their growth.  Once the
surrounding cocoon of jets reaches a certain limit, it probably becomes
Rayleigh--Taylor unstable and might frustrate the growth of the jet.
Evidently, additional physical phenomena (e.g. magnetic fields) play a
role in the formation and evolution of astrophysical jets, and should
be considered in a more detailed description.

\section{Acknowledgements}
  We would like to thank Tigran Arshakian, Jorge Cant\'o, Manoj
Choudhury and Enrico Ramirez--Ruiz for many fruitful discussions on \(
\mu \)--quasars, \( \gamma \)--ray bursts and jets in general while
preparing this letter.  We very much thank the comments made by
an anonymous referee, which gave a better physical understanding to
the problem.  S. Mendoza gratefully acknowledges financial support from
CONACyT (41443) and DGAPA--UNAM (IN119203).

\bibliographystyle{apj}
\bibliography{jet}

\label{lastpage}
\end{document}